\documentclass[twocolumn,pra,showpacs,preprintnumbers]{revtex4}
\usepackage{amsmath,amssymb}
\usepackage{graphicx}
\usepackage{dcolumn}
\usepackage{bm}

\newcommand{\hH}{{\mathcal{H}}}

\newcommand{\cro}[1]{\hat{#1}^{\dagger}}
\newcommand{\ano}[1]{\hat{#1}^{\phantom{\dagger}}}
\newcommand{\nhat}{{\hat{n}}}

\usepackage{epstopdf}

\begin{document}
\author{Santiago F. Caballero-Ben\'\i tez and Elena A. Ostrovskaya} 
\affiliation{
ARC Centre of Excellence for Quantum-Atom Optics and Nonlinear Physics Centre, Research School of Physics and Engineering, Australian National 
University, Canberra ACT 0200, Australia
}


\title{Ground state properties of a three-site Bose-Fermi ring with a small number of atoms}
\begin{abstract}
We investigate a three-site ring system with a small number of quantum degenerate bosons and fermions. By means of the exact diagonalization of the Bose-Fermi-Hubbard Hamiltonian, we show that the symmetry of the ground state configuration is a function of both the boson-boson and the inter-species interaction in the system.  The phase diagram of the system, constructed by computing the exact two-body spatial correlations, reveals nontrivial insulating phases that exist even in the strong bosonic tunneling limit and for incommensurate filling of bosons.  These insulating phases are due to the inter-species interactions in the system and are not necessarily accompanied by the suppression of the particle number fluctuations.   
\end{abstract}
\maketitle

\section{Introduction}

The ability to load ultracold quantum degenerate gases into periodic or quasi-periodic potentials of different geometries, created by optical lattices \cite{Bloch} has opened the way to realization of well known condensed matter systems with atoms of fermionic  \cite{FH-exp} and bosonic character \cite{BH-exp}. Such model systems allow us to gain deeper understanding of the fundamental problems in many body physics beyond the weak coupling, mean-field regime, such as transition between superfluid and Mott-insulator \cite{review}. In the strongly interacting regime, new phenomena can emerge, such as quantum magnetism \cite{spin_liq} and the supersolid phase \cite{supersolid}. In general, these emergent phenomena belong to the class of quantum phase transitions and their mechanisms are strongly dependent both on the geometry of the lattice and the interaction between the atoms loaded into the lattice. The theoretical and experimental studies of these effects in large many-sites lattice systems are important both from the fundamental and applied points of view. For example, the physics of ultracold lattice systems where frustration can occur, is fundamental for the development of fault tolerant systems and is important for quantum computing \cite{quant_comp}. 

On the other hand, the possibility to address atoms in single sites \cite{exp_cont_sites} and  to create mixtures of bosons and fermions \cite{exp_mix} draws the attention to the study of finite, inhomogeneous systems \cite{BF_inhom} with complex inter-species interactions. The study of such two- or three-dimensional systems enables understanding of basic physics of more complicated systems, such as interaction-dependent properties of ground states (such as frustrations) and the role of symmetry breaking in transitions between different quantum phases. Small, strongly correlated systems may also lend themselves to the process of controlled quantum state preparation and manipulation making it potentially useful in quantum information, quantum-limited measurements, and atomtronics \cite{santos_threewell,atomtronics}.

In this paper we consider a minimal finite two-dimensional lattice model, namely a three-site ring of a Bose-Fermi ultracold mixture. Such a three-site system may be realized experimentally by engineering magnetic microtraps on an atomic chip, or by combining a harmonic potential with a triangular or Kagome lattice, as suggested in \cite{CLee}. With the small number of atoms, this system lends itself to the Bose-Fermi-Hubbard model solvable by means of direct diagonalization.  We consider the ground state of the system and investigate how the admixture of fermions leads to various phases, depending on the filling factor and inter-species interaction strength. In particular, we consider unusual insulating phases resulting from the inter-species interaction, which is connected to the existence of macroscopic self-trapping states in the mean-field regime \cite{santiago_bf}.

\section{The model and the ground state configuration} 
Based on the standard Bose-Fermi Hubbard model (see, e.g., \cite{BF_inhom}), the Hamiltonian for the three-site ring can be written as follows: 
\begin{equation}
\label{Ham}
\hH=\hH_{b}+\hH_{f}+\hH_{bb}+\hH_{bf},
\end{equation}
where,
\begin{eqnarray}
\hH_{\xi}&=&-T_\xi\sum_{<l,m>}\left(\cro{\xi}_l\ano{\xi}_m+\cro{\xi}_m\ano{\xi}_l\right),
\\
\hH_{bb}&=&\frac{U_{bb}}{2}\sum_{l=1}^3 \nhat^{b}_l\left(\nhat^{b}_l-1\right),
\\
\hH_{bf}&=&U_{bf}\sum_{l=1}^3 \nhat^{b}_l \nhat^{f}_l.
\end{eqnarray}
Here $\cro{b}$ ($\ano{b}$) are the creation (annihilation) operators for the bosons and $\cro{f}$ ($\ano{f}$) the creation (annihilation) operators for the fermions; the number operators are: $\nhat^{\xi}=\cro{\xi}\ano{\xi}$, $\xi\in\{b,f\}$. The nearest neighbour tunneling coefficient is $T_\xi$, the intra- and inter-species interaction strengths are $U_{bb}>0$ and $U_{bf}$, respectively. 

In general, one can write a state vector of the system as follows:
$
|\tilde{\Psi}\rangle=|\tilde{\Psi}\rangle_f\otimes|\tilde{\Psi}\rangle_b=|n_1^f,n_2^f,n_3^f;\rangle_f\otimes|n^b_1,n^b_2,n^b_3\rangle_b,
$
where, $n^f_1+n^f_2+n^f_3=N_f$, and $n^b_1+n^b_2+n^b_3=N_b$.

\begin{figure} 
\includegraphics[width=\columnwidth]{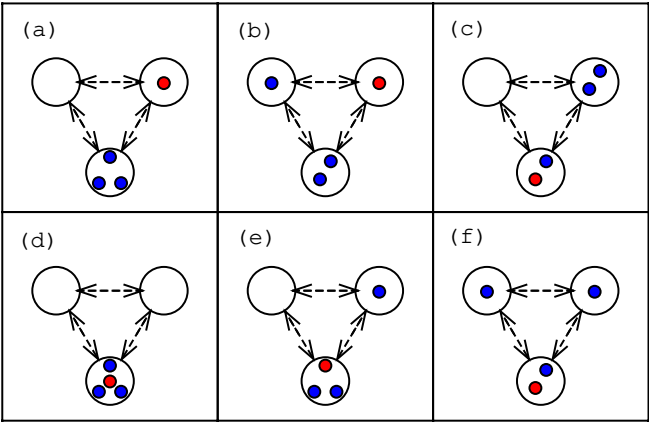} 
\caption{(Color online) Schematics of the on-site state configurations of three bosons (blue circles) and one fermion (red circle), with the lowest energy. The figures correspond to the states: (a) $|1\rangle_+$, (b) $|2\rangle_+$, (c) $|3\rangle_+$, and (f) $|4\rangle_+$ for repulsive inter-species interaction, and (d)  $|1\rangle_-$, (e) $|2\rangle_-$, (c) $|3\rangle_-$, and (f) $|4\rangle_-$ for inter-species attraction. } 
\label{states}
\end{figure}
\begin{figure} 
  \includegraphics[width=\columnwidth]{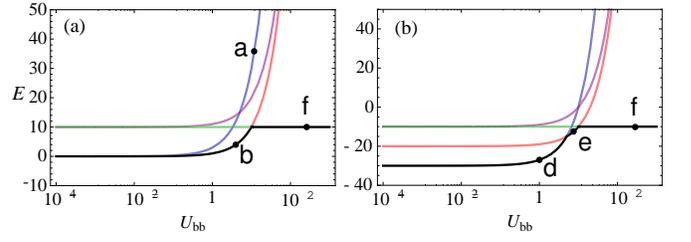}
\caption{(Color online) The energy of the states with the largest contribution to the ground state as a function of $U_{bb}$ for inter-species (a) repulsion and (b) attraction with the fixed magnitude $|U_{bf}|=10$. States with different symmetries are marked with the points. The unmarked states in (a) and (b) are  $|3\rangle_+$ and $|3\rangle_-$, respectively. } 
\label{energy}
\end{figure}

In the case of inter-species repulsion and $N_f=1$, the states corresponding to the lowest energy levels are the ones with the following on-site particle distributions:
\begin{eqnarray*}
|1\rangle_+&\rightarrow&|1,0,0\rangle_f\otimes| 0,3,0\rangle_b,
\\
|2\rangle_+&\rightarrow&|1,0,0\rangle_f\otimes|0,1,2\rangle_b,
\\
|3\rangle_+&\rightarrow&|0,0,1\rangle_f\otimes| 0,2,1\rangle_b,
\\
|4\rangle_+&\rightarrow&|1,0,0\rangle_f\otimes| 1,1,1\rangle_b,
\end{eqnarray*}
where $|1\rangle_+$, $|2\rangle_+$ and  $|3\rangle_+$ have six-fold degeneracy, while $|4\rangle_+$ is a triplet. These states are schematically shown in Fig. \ref{states} and the corresponding energies are plotted in Fig. \ref{energy} (a).

\begin{figure} 
\includegraphics[width=\columnwidth]{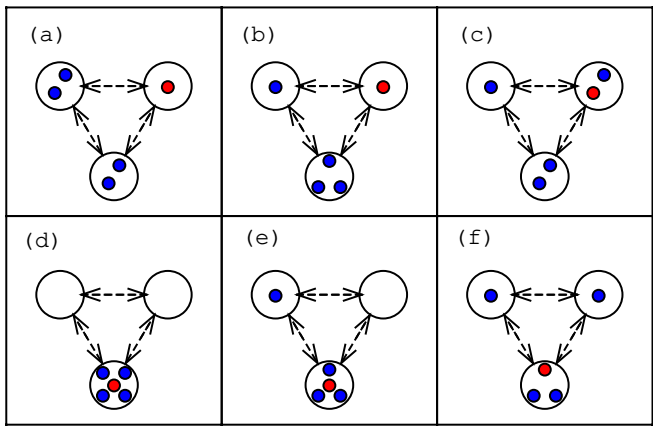} 
\caption{(Color online) Schematics of the on-site state configurations of four bosons (blue circles) and one fermion (red circle), contributing to the lowest energy band. The panels (a), (b), (c), and (f) correspond to repulsive inter-species interaction and (d), (e), and (f) -- to inter-species attraction. } 
\label{states_inc}
\end{figure}

For attraction and $N_f=1$ (see Fig. \ref{states} and Fig. \ref{energy} (b)), the lowest-lying energy states are as follows:
\begin{eqnarray*}
|1\rangle_-&\rightarrow&|0,0,1\rangle_f\otimes| 0,0,3\rangle_b,
\\
|2\rangle_-&\rightarrow&|1,0,0\rangle_f\otimes| 2,0,1\rangle_b,
\\
|3\rangle_-&\rightarrow&|1,0,0\rangle_f\otimes| 1,0,2\rangle_b,
\\
|4\rangle_-&\rightarrow&|0,1,0\rangle_f\otimes| 1,1,1\rangle_b,
\end{eqnarray*}
where the states with symmetry $|1\rangle_-$ and $|4\rangle_-$ are triplets and $|2\rangle_-$ and $|3\rangle_-$ have six-fold degeneracy. Similar ground state configurations occur at incommensurate filling, as shown in Fig. \ref{states_inc}. 

\begin{figure} 
  \includegraphics[width=\columnwidth]{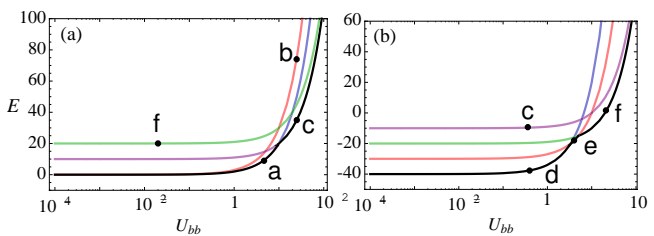}
\caption{(Color online) The energy of the states with the largest contribution to the ground state as a function of $U_{bb}$ for inter-species (a) repulsion and (b) attraction with the fixed magnitude of inter-species interaction $|U_{bf}|=10$. Energies of states with different symmetries shown in Fig. \ref{states_inc} are marked with the corresponding letters. } 
\label{energy_inc}
\end{figure}

As the bosonic interaction strength changes, the ground state of the system is changing symmetry. The hierarchy of the lowest lying energy levels corresponding to different states is shown in Fig. \ref{energy} for the commensurate and in Fig.  \ref{energy_inc} for incommensurate filling of bosons. In the case of commensurate filling, for a fixed value of the inter-species interaction and growing $U_{bb}$, the ground state structure evolves from a mixture of degenerate states $(a)$ and $(b)$ to the state $(f)$ (in Fig \ref{states}) in the case of inter-species repulsion, and from  $(d)$ to $(e)$ and $(f)$ in the case of inter-species attraction. Similarly, in the case of incommensurate filling, the ground state evolves from a mixture of states $(a)$ and $(b)$ to $(c)$ and to the mixture of degenerate states $(c)$ and $(f)$ (see Fig. \ref{states_inc}) for inter-species repulsion, and from $(d)$ to $(e)$, to the mixture of states $(c)$ and $(f)$ for the attractive inter-species interaction.

\begin{figure}
  \includegraphics[width=\columnwidth]{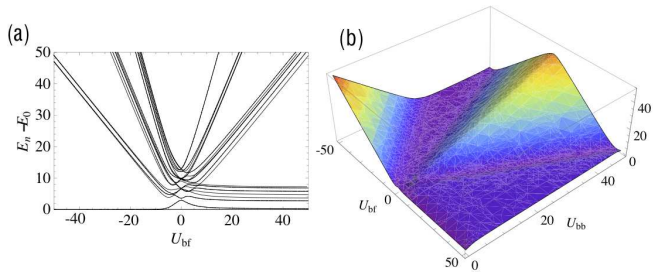}
\caption{(Color online) (a) Structure of the energy spectrum vs. $U_{bf}$ for commensurate filling of bosons ($U_{bb}=1$, $N_b=3$). (b) The dependence of the gap, $\Delta=E_1-E_0$ on the intra- and inter-species interaction strengths for $N_b=3$. }
\label{gap}
\end{figure}

The energy spectrum of this small scale system exhibits a gap, $\Delta$, given by the energy difference between the ground state and first excited manifold [see Fig. \ref{gap} (a)]. For the commensurate filling of bosons this gap opens up at sufficiently strong boson repulsion for both repulsive and attractive inter-species interaction, as well as for $U_{fb}=0$ [see Fig. \ref{gap}(b)], and its magnitude is proportional to the interaction strength. It is reminiscent of the gap in the excitation spectrum of bosonic systems in lattices that indicates superfluid (SF) to Mott-insulator (MI) transition. Indeed, using the exact diagonalization of the Hamiltonian (\ref{Ham}) and extracting characteristic behavior of tunneling correlations and particle number fluctuations, we can construct the phase diagram of the ground state and examine the transition to the insulating states in our small-scale Bose-Fermi system.
 
\section{Phase diagram}
The typical characteristic quantity for different quantum phases of an ultracold quantum degenerate gas in a the lattice system is the spatial (tunneling) correlation for the bosonic component $\eta_b$  and the tunneling correlation of the fermionic component $\eta_f$, given by $\eta_\xi=|\langle \cro{\xi}_{i}\ano{\xi}_{i+1}\rangle|/n_{\textrm{avg}}^\xi$, $\xi\in\{b,f\}$,  where $n_{\textrm{avg}}^f=N_f/3$. The boson tunneling correlation can be used to construct the phase diagram of the model, depending on the interactions. In a bosonic lattice this quantity tends to zero when the system is in the Mott-insulator regime, and approaches one when the system is in the superfluid regime.  Similarly, in our small, finite, mixed-species system, this quantity can be used to delineate between an insulating and a superfluid state. The fermion tunneling correlation measures the mobility of the fermion in the system. When its value is close to one, the fermion is mobile, and when it is zero, the fermion is pinned to one of the sites. 

\subsection{Commensurate boson filling}
\begin{figure}
  \includegraphics[width=\columnwidth]{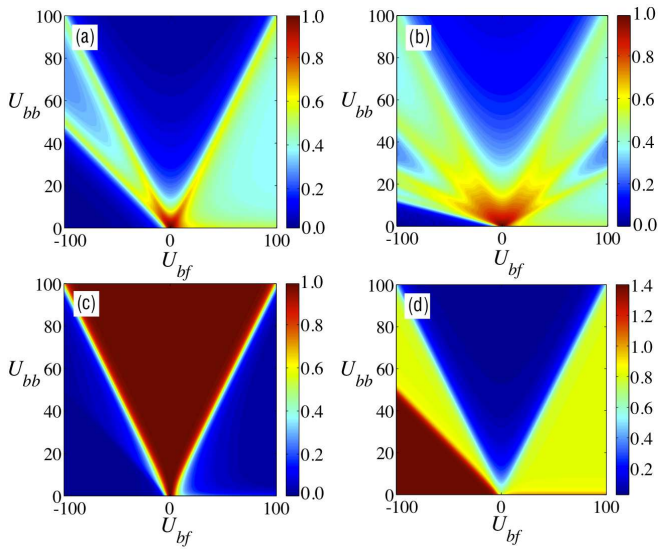}
\caption{(Color online) (a,b) Density plots of the boson tunneling correlation, $\eta_b(U_{bf},U_{bb})$, for (a) $N_b=3$ and (b) $N_b=9$. (c,d) Density plots of the (c) fermion tunneling correlation, $\eta_f(U_{bf},U_{bb})$ and (d) boson number fluctuations, $\sigma(U_{bf},U_{bb})$ for $N_b=3$.}
\label{phase_com}
\end{figure}

As a start, we analyze the components of the ground state for the system with a commensurate number of bosons, $N_f=1$  ($n_{\textrm{avg}}^f=1/3$), and for the fixed inter-species interaction strength $|U_{bf}|=10$. The boson tunneling correlation for $N_b=3,9$ is shown in Fig. \ref{phase_com}.  Due to the small number of particles, there exists a distinct crossover region between the SF ($\eta_b=1$) and insulating ($\eta_b=0$) regimes \cite{CLee}, where $0<\eta_b<1$. When this quantity is closer to either of the two extreme values, we will refer to it as a SF- or MI-enhanced regime, respectively. It can be seen that the system exhibits a rich phase diagram with an asymmetric behavior depending on the sign of the inter-species interaction and the number of bosons. 

For the case of attraction between bosons and fermions, $U_{bf}<0$, a new insulating phase corresponding to the ground state $|1\rangle_-$ (Fig. \ref{states}, d) appears. It is strikingly different from the regular MI-like insulating state in the pure bosonic system, $|4\rangle_{\pm}$ (see Fig. \ref{states}), which appears at $U_{bf}=0$ and dominates the phase diagram for larger inter-species interaction strength. In this interaction-induced insulating phase the bosonic occupation of the ring sites is strongly unbalanced, in analogy with the mean-field macroscopic self-trapped states of a Bose-Fermi mixture \cite{santiago_bf}. As one increases the repulsion between bosons, a new  SF enhancement region appears and then the ground state of the system is once again dominated by the regular insulating state $|4\rangle_{-}$. With the increasing inter-species interaction and the number of bosons, the SF enhancement region breaks into several filaments separated by the insulating states $|2\rangle_-$ that appear due to interaction between bosons and fermions.  This change in the structure of ground state can also be followed in Fig. \ref{energy} (b).

In the case of inter-species repulsion, $U_{bf}>0$,  the system is superfluid for small $|U_{bf}|$. In Fig. \ref{energy}(a) one can clearly see that the ground state in this region is superposition of degenerate states $|1\rangle_{+}$ and $|2\rangle_{+}$, which is a typical sign of frustration. Insulating regions appear for larger inter-species interaction and are dominated by the $|2\rangle_{+}$ state [see Fig. \ref{phase_com} (b)]. For large $U_{bb}$ transition to the regular insulating state  $|4\rangle_+$ occurs.  

For both signs of inter-species interaction, the behavior in the regions where $U_{bb}\sim U_{bf}$ is SF-enhanced. The effect of increasing the number of bosons is to scale up the regions of SF behaviour and introduce additional insulating regions due to inter-particle interactions. The appearance of the gap in the energy spectrum (Fig.  \ref{gap}) correlates exactly with the insulating regions in the phase diagram.

While the phase diagrams in Fig. \ref{phase_com} (a,b) are based on the behaviour of the bosonic fraction, it is useful to examine the tunneling correlation of the fermion in the system. As seen in Fig. \ref{phase_com}(c), in the region corresponding to the regular bosonic insulating state, (i.e.  for $|U_{bb}|\gg 1$) and for low inter-species coupling, the fermions are free to hop between the ring sites, which  can also be deduced from the symmetry of the $|4\rangle_\pm$ state [see Fig. \ref{states}]. In the mean-field picture \cite{santiago_bf} this behavior reflects the fact that the effective interaction-induced potential seen by the fermion is weak and completely symmetric. As one increases $|U_{bf}|$, the fermions localize and no tunneling is possible. In contrast to the regular insulating phase, the bosonic insulating phases that arise purely due to the inter-species interaction and correspond to symmetry-broken states (e.g., $|1\rangle_-$ or $|2\rangle_+$), naturally give rise to the regions of suppressed tunneling of the fermion. 

To characterize the interaction-induced insulating phases further, we look at the boson number fluctuations in the system given by $\sigma=\sqrt{\langle\hat{n}_i^b\hat{n}_i^b\rangle-\langle\hat{n}_i^b\rangle^2}/n^b_{avg}$, see Fig.\ref{phase_com}(d). As expected from the MI behavior, the fluctuations in the regular insulating phase (with the $|4\rangle_{\pm}$ symmetry) approach zero as one increases the strength of repulsion between bosons. In contrast, for attractive inter-species interaction the interaction-induced insulating region at low values of $U_{bb}$ is dominated by fluctuations, which can be taken as a signature of the new insulating state.

\subsection{Incommensurate boson filling}

\begin{figure}
  \includegraphics[width=\columnwidth]{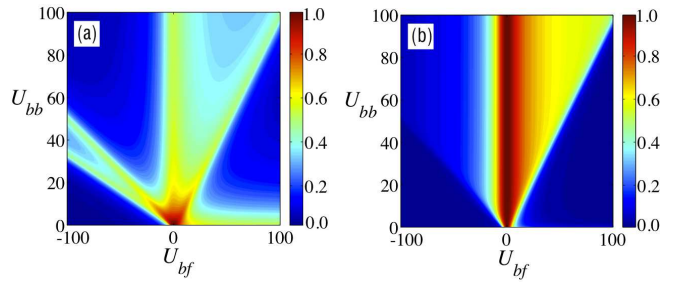}
\caption{(Color online) Density plot of the (a) boson and (b) fermion tunneling correlation for $N_f=1$,$N_b=4$. }
\label{phase_inc}
\end{figure}

\begin{figure}
\includegraphics[width=\columnwidth]{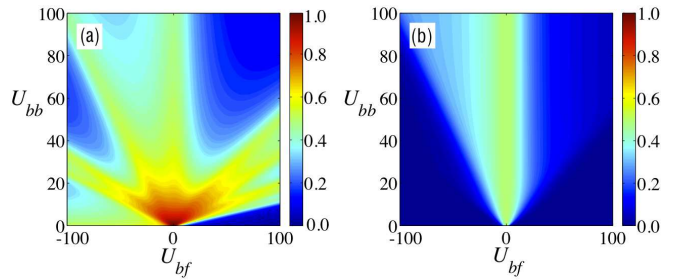}
\caption{(Color online)  Density plot of the (a) boson and (b) fermion tunneling correlation for $N_f=2$,$N_b=10$. }
\label{2fermions}
\end{figure}

It has been noted (see, e.g., \cite{CLee,finite_lattice}) that for a few-particle pure bosonic system at incommensurate filling a small superfluid fraction is always present, therefore no insulating state can occur. We find that this is indeed the case in our system with no fermions (or $U_{bf}=0$). The insulating phases in this case occur only in the presence of fermions, with non-zero inter-species interaction strength. For the repulsive inter-species interaction the insulating phase at large $U_{bb}$ is suppressed, as shown in Fig.  \ref{phase_inc}(a), which is typical for $N_b=3m+1$ bosons with $m$ being a positive integer.  In contrary, the interaction-induced insulating region is very prominent for small $N_b$ and is broken up by regions of enhanced superfluidity for larger $N_b$. For inter-species attraction both the regular MI-like insulating region at large $|U_{bb}|$ and the insulating region at lower values of $U_{bb}$ are retained due to the inter-species interaction.  As in the case of commensurate filling, at large $U_{bb}$ the new insulating domains that appear due to the inter-species interaction are dominated by the ground state configuration $(f)$ in Fig. \ref{states_inc}. In general, both for incommensurate and commensurate filling of bosons the boundaries of the insulating regions are pushed to larger interaction strength values as one increases the number of bosons. 

In striking difference from the commensurate case, the tunneling of fermions is almost completely suppressed in the attractive inter-species interaction region, $U_{bf}<0$ [see Fig.  \ref{phase_inc}(b)]. For repulsive inter-species interaction, $U_{bf}>0$, the fermion gains mobility as one increases the boson interaction strength, $U_{bb}$. This is due to the fact that, for larger $U_{bb}$ and repulsive inter-species interaction, the insulating phase is dominated by the state $(c)$ in Fig. \ref{states_inc} [see also Fig. Fig. \ref{energy_inc} (a)]. In this case the fermion is always able to hop between the sites with lower bosonic occupation numbers. In contrast, in the case of inter-species attraction the insulating phase is dominated by the state  $(f)$ in Fig. \ref{states_inc} [see also Fig. \ref{energy_inc} (b)] and the fermions are pinned to a site with the highest occupation of bosons. For weaker inter-species interaction, $U_{bf}<U_{bb}$, the fermion is delocalized as in the commensurate case. 

\subsection{Role of the fermion filling factor}

So far, we have considered one fermion interacting with $N_b$ bosons. Therefore, the results presented above are also applicable to a mixture of two bosonic species (see, e.g., \cite{dw_bosons}) with a single atom in one of the components. In the case of $N_f\neq1$, the Bose-Fermi-Hubbard model is still valid for the fermion filling factor less than $1$ (the system with no fermions is equivalent to the system with three fermions), and the fermion statistics influences the ground state configuration. In particular, the system is particle-hole symmetric for $1/3$ and $2/3$ filling of fermions. This fact is reflected in the behavior of the phase diagram, so that  the case of $1/3$ fermion filling with repulsive inter-species interactions corresponds to the case of $2/3$ fermion filling with attractive inter-species interaction. For example, in the case of $2/3$ filling of fermions the interaction-induced insulating regions appear for repulsive rather then attractive interaction between species, as seen in Fig. \ref{2fermions}. Other characteristic properties of the system, such as the behaviour of the fermion tunneling correlations and bosonic number fluctuations, are qualitatively the same as for $N_f=1$.

\section{Conclusions}
In conclusion, we have analyzed the ground state of a small-scale system of quantum degenerate bosons and fermions in a three-site ring configuration. We have restricted the consideration to the fermion filling factor less or equal than one, which as allowed us to employ a standard Bose-Fermi-Hubbard Hamiltonian. By examining the tunneling correlations and particle fluctuations in the system, we have found that the system admits mobile and insulating states that are analogous to the superfluid and Mott-insulator states in infinite lattices. The novel insulating states identified in this small-scale system for both commensurate and incommensurate filling of bosons, are purely due to the inter-species interactions, and can be controlled by controlling the interaction strengths and the number of fermions injected into the system.

This work is supported by the Australian Research Council (ARC). The authors acknowledge useful discussions with Dr. Chaohong Lee and Dr. Tristram Alexander.

 \end{document}